\documentclass[aps,pra,reprint,aps,twocolumn,superscriptaddress]{revtex4}

\usepackage{amsmath,amsthm,amssymb,graphicx,xcolor,float}

\usepackage{esint}
\usepackage[unicode=true]{hyperref}
\hypersetup{
     colorlinks=true,       		
     linkcolor=blue,          	
     citecolor=red,            
     urlcolor=magenta,           	
 }

\usepackage{mathrsfs,mathtools,mathdots,dsfont}
\usepackage{listings}
\usepackage{subfigure}

\newtheorem*{theorem*}{Theorem}

\newtheorem*{corollary*}{Corollary}

\newtheorem*{lemma*}{Lemma}

\newtheorem*{proposition*}{Example*}

\newtheorem*{conjecture*}{Conjecture}
\theoremstyle{definition}

\newtheorem*{definition*}{Definition}
\theoremstyle{remark}
\newtheorem{remark}{Remark}
\newtheorem*{remark*}{Remark}

\newcommand{\ket}[1]{\left|#1\right\rangle}
\newcommand{\bra}[1]{\left\langle#1\right|}

\begin{document}



\title{Peres-type Criterion of Einstein-Podolsky-Rosen Steering for Two Qubits}

\author{Yu-Xuan Zhang}
\affiliation{School of Physics, Nankai University, Tianjin 300071, People's Republic of China}

\author{Jing-Ling Chen}
\email{chenjl@nankai.edu.cn}
\affiliation{Theoretical Physics Division, Chern Institute of Mathematics, Nankai University, Tianjin 300071, People's Republic of
	 	China}

\date{\today}

\begin{abstract}
Quantum nonlocality manifests in multipartite systems through entanglement, Bell's nonlocality, and Einstein-Podolsky-Rosen (EPR) steering. While Peres's positive-partial-transpose criterion provides a simple and powerful test for entanglement, a comparably elegant spectral criterion for detecting EPR steering remains an open challenge. In this work, we systematically explore whether a Peres-type criterion can be established for EPR steering in the two-qubit system. Focusing on rank-2 (including rank-1) states and the two-qubit Werner state, we analyze the eigenvalues of their partially transposed density matrices and construct a significant steering criterion based on symmetric combinations of these eigenvalues. We prove that this criterion serves as a necessary and sufficient condition for steerability for the Werner state, all two-qubit pure states, all two-qubit rank-2 states. Furthermore, we validate the criterion for higher-rank states (rank-3 and rank-4) and show that the results align with known steering inequalities. Our findings suggest a more unified framework for detecting quantum nonlocality via partial transposition and open avenues for further theoretical and numerical investigations into steering detection.
\end{abstract}

\maketitle


\section{Introduction}

Quantum nonlocality manifests in multipartite quantum systems through various forms, ranging from quantum entanglement (QE) \cite{Einstein1935,Schrodinger35,RMP2009} to Bell's nonlocality \cite{Bell,RMP2014} and Einstein-Podolsky-Rosen (EPR) steering \cite{2017RPPCavalcanti,RMP2020}. Among these, EPR steering -- a distinct nonlocal phenomenon intermediate between quantum entanglement and Bell nonlocality -- describes the ability of one observer, through local measurements, to affect the state of a distant particle, in a way that cannot be explained by the local-hidden-state (LHS) models. Since its rigorous formalization by Wiseman, Jones, and Doherty in 2007 \cite{WJD07,WJD07PRA}, EPR steering has been recognized not only as a fundamental feature of quantum mechanics, but also as a key resource for one-sided device-independent quantum information tasks, such as quantum key distribution and subchannel discrimination \cite{Nielsen2000}.

For quantum entanglement, the positive-partial-transpose (PPT) criterion, introduced by Peres in 1996 \cite{AP1996}, provides a simple yet powerful necessary condition for separability. Remarkably, for the two-qubit and qubit-qutrit systems, Peres's criterion is also a necessary and sufficient condition. This criterion has profoundly influenced entanglement detection and quantification, offering a clear operational tool based on the eigenvalues of the partially transposed density matrix. In contrast, the detection of EPR steering still relies largely on the inequality-based criteria or semi-definite programming methods, which often lack the simplicity and algebraic clarity of Peres's criterion. Despite significant advances in characterizing steerability, a comparably elegant and universally applicable spectral criterion -- akin to the PPT test -- remains an open challenge. Several works have proposed conjectured steering criteria expressed in terms of the eigenvalues of the partially transposed state, yet these lack rigorous proofs or fail to be generally valid \cite{Chen2011}.

This work aims to bridge this gap by establishing a Peres-type spectral criterion for EPR steering, which carries significant scientific value in several aspects:
First, it provides an eigenvalue-based detection tool that retains the algebraic simplicity and conceptual clarity of Peres's original criterion, thereby facilitating both analytical treatments and numerical implementations.
Second, by extending the PPT formalism from entanglement to steering, our approach promotes a more unified spectral perspective on quantum nonlocality, linking different nonclassical correlations through the same mathematical operation -- partial transposition.
Third, the criterion is proved to be necessary and sufficient for important families of states (the Werner state \cite{werner89}, all pure states, and all rank-2 states), offering a rigorous and efficient detection method for these physically relevant cases.
Finally, the proposed criterion respects the known hierarchy among entanglement, steering, and Bell nonlocality, and its performance on higher-rank states aligns with existing steering inequalities, confirming its consistency and broader applicability.

In this work, we systematically investigate whether whether a Peres-type criterion can be established for EPR steering in a two-qubit system. We focus particularly on the rank-2 (including rank-1 as a special case) quantum states -- a broad and physically relevant family -- and study the eigenvalues of their partial transpose. Through careful analytical and algebraic examination, we propose a Peres-type criterion that reflects the underlying symmetry and the steerability structure of quantum states. We find that, for all the rank-2 (and rank-1) states and the two-qubit Werner state \cite{werner89}, the steering analogue of the PPT criterion serves as a necessary and sufficient to detect steerability. We also examine other rank-3 and rank-4 quantum states using this Peres-type criterion, and the results are consistent with known steering inequalities. Our results point toward a more unified framework for detecting quantum nonlocality through partial transposition.

The paper is organized as follows. In Sec. \ref{sec:peres1}, we brief review Peres's criterion for quantum entanglement in a two-qubit system. In Sec. \ref{sec:peres2}, we brief review an alternative Peres-type criterion for EPR steering given in Ref. \cite{Chen2011}, and analyze its advantage and deficiency of detecting the steerability. In Sec. \ref{sec:peres3}, we construct a new Peres-type criterion, which serves as a necessary and sufficient condition to detect the steerability for the two-qubit Werner state, all two-qubit pure states, all two-qubit rank-2 states.
In Sec. \ref{sec:peres4}, we prove the criterion for rank-2 states. In Sec. \ref{sec:peres5}, we test the criterion on two well-known rank-3 and rank-4 states and find reasonable agreement. In the final section, we present conclusions and an outlook for future work.


\section{Brief Review of Peres's Criterion}\label{sec:peres1}

In this section, let us we make a brief review of Peres's criterion for quantum entanglement. For simplicity, we restrict to the two-qubit system.

As it is well-known, a qubit system is described by the $2\times 2$
 density matrix as
 \begin{eqnarray}
&& \rho = \frac{1}{2}\left(\openone+ {\vec \sigma} \cdot \vec{r}\right),
 \end{eqnarray}
where $\openone$ is the $2\times 2$ identity matrix, $\vec{r}$ is the Bloch vector with $|\vec{r}| \le 1$ (here $|\vec{r}|=1$ corresponds to pure states and $|\vec{r}|<1$ corresponds to mixed states), ${\vec \sigma}=(\sigma_1, \sigma_2, \sigma_3)$ is the vector of Pauli matrices with
\begin{eqnarray}
 \label{S-3}
   \sigma_x=
  \left(
    \begin{array}{cc}
      0 & 1 \\
      1 & 0 \\
    \end{array}
  \right),\; \sigma_y=
  \left(
    \begin{array}{cc}
      0 & -{\rm i} \\
      {\rm i} & 0 \\
    \end{array}
  \right),\;  \sigma_z=
  \left(
    \begin{array}{cc}
      1 & 0 \\
      0 & -1 \\
    \end{array}
  \right).
   \end{eqnarray}
Similarly, for a two-qubit system, a quantum state is completely described by the following $4 \times 4$ density matrix
  \begin{equation}\label{eq:state-1}
   \rho_{AB} = \frac{1}{4} ( \openone \otimes \openone +
   {\vec \sigma}^A \cdot \vec{u} \otimes \openone+
   \openone \otimes {\vec \sigma}^B \cdot \vec{v}+
   \sum_{i,j=1}^3 \beta_{ij} \sigma_i^A \otimes \sigma_j^B ).
 \end{equation}
Let Alice and Bob share the state $\rho_{AB}$, one can obtain the two reduced density matrices for Alice and Bob as
 \begin{eqnarray}
&&   \rho_A^{\phantom{2}} = {\rm tr}_B (\rho_{AB}^{\phantom{2}})=
\frac{1}{2} \left(\openone+ {\vec \sigma}^A \cdot \vec{ u}\right),\nonumber\\
&&   \rho_B^{\phantom{2}} = {\rm tr}_A (\rho_{AB}^{\phantom{2}})
= \frac{1}{2} \left(\openone+ {\vec \sigma}^B \cdot \vec{v}\right),
\end{eqnarray}
 here $\vec{u}$ and $\vec{v}$ are respectively the Bloch vectors for Alice and Bob, and $\beta_{ij}$ are some real numbers. If ${\rm tr} [\rho_{AB}^2]=1$, the corresponding state is pure, while if ${\rm tr} [\rho_{AB}^2] <1$, the corresponding state is mixed.

Now we consider the partial transpose of Bob's system. The partial transposed density matrix is denoted by $\rho_{AB}^{\rm T_B}$, where ``${\rm T}$'' represents the transposition of a matrix. There are two methods to express the matrix $\rho_{AB}^{\rm T_B}$. The first method is based on Eq. (\ref{eq:state-1}), namely, we have
  \begin{eqnarray}
  \rho_{AB}^{\rm T_B} &=& \frac{1}{4} \biggr[\openone \otimes \openone +
   {\vec \sigma}^A \cdot \vec{u} \otimes \openone+
   \openone \otimes \left({\vec \sigma}^B\right)^{\rm T} \cdot \vec{v}\nonumber\\
  && +\sum_{i,j=1}^3 \beta_{ij} \sigma_i^A \otimes \left( \sigma_j^B \right)^{\rm T} \biggr],
 \end{eqnarray}
with
  \begin{equation}
\left({\vec \sigma}^B\right)^{\rm T} = (\sigma_x^{\rm T}, \sigma_y^{\rm T}, \sigma_z^{\rm T})
=(\sigma_x, -\sigma_y, \sigma_z).
 \end{equation}
The second method is as follows. If the $4\times 4$ density matrix reads
\begin{eqnarray}
 \rho_{AB}=\left(
  \begin{array}{cccc}
 \rho_{11} & {\rho_{12}} & \rho_{13} & {\rho_{14}} \\
 {\rho_{21}} & \rho_{22} & {\rho_{23}} & \rho_{24} \\
 \rho_{31} & {\rho_{32}} & \rho_{33} & {\rho_{34}} \\
 {\rho_{41}} & \rho_{42} & {\rho_{43}} & \rho_{44} \\
  \end{array}
\right),
\end{eqnarray}
then the partial transposed density matrix is given by
\begin{eqnarray}\label{eq:trans}
 \rho_{AB}^{\rm T_B}=\left(
  \begin{array}{cccc}
 \rho_{11} & {\rho_{21}}  & \rho_{13} & {\rho_{23}} \\
 {\rho_{12}} & \rho_{22} & {\rho_{14}} & \rho_{24} \\
 \rho_{31} & {\rho_{41}} & \rho_{33} & {\rho_{43}} \\
 {\rho_{32}} & \rho_{42} & {\rho_{34}} & \rho_{44} \\
  \end{array}
\right).
\end{eqnarray}

Let us study the eigen-equation of $\rho^{T_B} $, i.e.,
\begin{eqnarray}
\det[\lambda\, \mathbb{I}-\rho^{T_B} ]=0,
\end{eqnarray}
where $\mathbb{I}=\openone\times \openone$ is the $4\times 4$ identity matrix. Let $\{\lambda_1,\lambda_2,\lambda_3,\lambda_4\}$ be four eigenvalues of $\rho_{AB}^{T_B}$ (note that $\rho_{AB}^{T_B}$ and $\rho_{AB}^{T_A}$ share the same eigenvalues), then we easily have
\begin{eqnarray}
&& (\lambda-\lambda_1)(\lambda-\lambda_2)(\lambda-\lambda_3)(\lambda-\lambda_4)=0,
\end{eqnarray}
i.e.,
\begin{eqnarray}
&& \lambda^4-\Lambda_1\lambda^3+ \Lambda_2 \lambda^2 -\Lambda_3 \lambda +\Lambda_4=0,
\end{eqnarray}
with
\begin{eqnarray}
&& \Lambda_1=\lambda_1+\lambda_2+\lambda_3+\lambda_4, \nonumber\\
&& \Lambda_2=\lambda_1\lambda_2+\lambda_1\lambda_3+\lambda_1\lambda_4+\lambda_2\lambda_3+\lambda_2\lambda_4+\lambda_3\lambda_4, \nonumber\\
&& \Lambda_3=\lambda_1\lambda_2\lambda_3+\lambda_1\lambda_2\lambda_4+\lambda_1\lambda_3\lambda_4+\lambda_2\lambda_3\lambda_4, \nonumber\\
&& \Lambda_4=\lambda_1\lambda_2\lambda_3\lambda_4.
\end{eqnarray}
One may observe that $\Lambda_k$'s ($k=1, 2, 3, 4$) have a permutation symmetry under the permutation of $\lambda_i$ and $\lambda_j$. Due to ${\rm tr} [\rho_{AB}] =1$ one has
\begin{eqnarray}
&& \Lambda_1=\lambda_1+\lambda_2+\lambda_3+\lambda_4=1,
\end{eqnarray}
which means that these four $\lambda_k$'s are not independent, e.g., $\lambda_1=1-(\lambda_2+\lambda_3+\lambda_4)$.

Without loss of generality, let $\{\lambda_1,\lambda_2,\lambda_3,\lambda_4\}$ be in the small-to-large order
(i.e., $\lambda_1 \leq \lambda_2 \leq \lambda_3 \leq \lambda_4$), here $\lambda_1$ is the smallest eigenvalue. In 1996, through intuitive observation, Peres presented the PPT criterion as follows
\begin{eqnarray}\label{eq:criterion-e}
\mathcal {E}= \lambda_1 \geq 0,
\end{eqnarray}
which is a necessary condition for separability, i.e.,  a separable state must satisfy the criterion. However, for a two-qubit system, it is also a sufficient condition to detect entanglement. Explicitly, if the criterion (\ref{eq:criterion-e}) is violated, then the state $\rho_{AB}$ is entangled, otherwise, the state $\rho_{AB}$ is separable.

\begin{remark} \emph{Detecting the Werner state by Peres's criterion.} Peres's criterion is a highly efficient tool to figure out the critical value of entanglement for the two-qubit Werner state \cite{werner89}, here $V_{\rm cr}^{\rm QE}=1/3$ is the critical value between entanglement and separability. The two-qubit Werner is given by
\begin{eqnarray}
\rho_{\rm Werner}&=& V |\Psi\rangle\langle \Psi|+(1-V)\frac{\mathbb{I}}{4}\nonumber\\
&=&\left(\begin{matrix}
\frac{1+V}{4}&0&0&\frac{V}{2}\\
0&\frac{1-V}{4}&0&0\\
0&0&\frac{1-V}{4}&0\\
\frac{V}{2}&0&0&\frac{1+V}{4}\end{matrix} \right),
\end{eqnarray}
where $|\Psi\rangle=(|00\rangle+|11\rangle)/\sqrt{2}$ is the two-qubit maximal entangled state, and the parameter $V\in [0, 1]$. For the Werner state, the entangled region is $V\in(1/3,1]$. From Eq. (\ref{eq:trans}), one has the corresponding partial transposed density matrix as
\begin{eqnarray}
\rho_{\rm Werner}^{\rm T_B}
&=&\left(\begin{matrix}
\frac{1+V}{4}&0&0& 0\\
0&\frac{1-V}{4}& \frac{V}{2} &0\\
0& \frac{V}{2}&\frac{1-V}{4}&0\\
0 &0&0&\frac{1+V}{4}\end{matrix} \right),
\end{eqnarray}
whose four eigenstates are
\begin{eqnarray}\label{eq:werner1}
\lambda\in\left\{\frac{1-3V}{4},\frac{1+V}{4},\frac{1+V}{4},\frac{1+V}{4}\right\}.
\end{eqnarray}
Based on Peres's criterion, from
\begin{eqnarray}
\mathcal {E}= \lambda_1 = \frac{1-3V}{4} < 0,
\end{eqnarray}
we directly have the critical value $V_{\rm cr}^{\rm QE}=1/3$ and the entangled region is $V\in(1/3,1]$. $\blacksquare$
\end{remark}

\begin{remark} \emph{Detecting all pure entangled states by Peres's criterion.} Peres's criterion is also a powerful tool to detect all pure entangled states of two qubits. The two-qubit pure state is given by
\begin{eqnarray}
&& |\Psi(\theta)\rangle=\cos\theta|00\rangle+\sin\theta|11\rangle,
\end{eqnarray}
with $\theta\in[0,\pi/2]$. It has been known that the entangled region is $\theta\in (0,\pi/2)$, i.e., the pure state is separable only when $\theta=0$ or $\pi/2$. The density matrix form reads
\begin{eqnarray}
\rho_{\rm pure}&=&\left(\begin{matrix}
\cos^2\theta&0&0&\sin\theta\cos\theta\\
0&0&0&0\\
0&0&0&0\\
\sin\theta\cos\theta&0&0&\sin^2\theta\end{matrix} \right).
\end{eqnarray}
It is easy to work out the four eigenvalues of $\rho_{\rm pure}^{\rm T_B}$ as
\begin{eqnarray}\label{eq:pure-1}
&& \lambda\in\{-\sin\theta\cos\theta,\sin^2\theta ,\sin\theta\cos\theta,\cos^2\theta\}.
\end{eqnarray}
Based on Peres's criterion, from
\begin{eqnarray}
\mathcal {E}= \lambda_1 = -\sin\theta\cos\theta < 0,
\end{eqnarray}
we directly have the desired entangled region as $\theta\in (0,\pi/2)$. $\blacksquare$
\end{remark}

\begin{remark} \emph{A permutation-symmetry form of Peres's criterion.} Let us look at Eq. (\ref{eq:criterion-e}), one may notice that Peres's criterion uses only $\lambda_1$, thus lacking a permutation symmetry among four eigenvalues $\{\lambda_1,\lambda_2,\lambda_3,\lambda_4\}$.
Actually, Peres's criterion can be recast to the following permutation-symmetry form
\begin{eqnarray}
\mathcal {E}= \lambda_1 \lambda_2 \lambda_3 \lambda_4  \geq 0,
\end{eqnarray}
i.e.,
\begin{eqnarray}\label{eq:criterion-e-1}
\mathcal {E}= \Lambda_4  \geq 0,
\end{eqnarray}
due to the fact that for a two-qubit entangled state, only $\lambda_1<0$, and the other three eigenvalues are positive \cite{check1}.
The form (\ref{eq:criterion-e}) and the form (\ref{eq:criterion-e-1}) have their own merits. For the former, it is very simple when one adopts it to verify quantum entanglement (and the other three eigenvalues appear redundant); for the latter, its symmetric structures is very helpful for generalizing Peres' criterion from entanglement to steering, which we shall show in the below sections. $\blacksquare$
\end{remark}

\section{Brief Review of an Alternative Peres-type Criterion}\label{sec:peres2}

With the success of Peres's criterion for detecting quantum entanglement, it naturally gives rise to a scientific question: Is there a similar Peres-type criterion for EPR steering in a two-qubit system? The answer is yes. In 2011 (nearly five years after Wiseman \emph{et al.}'s work \cite{WJD07}), Chen, Su, Ye, Wu, and Oh (CSYWO) proposed a sufficient criterion to detect EPR steering of two-qubit density matrices \cite{Chen2011}. For the convenience of elaboration, we would like to call the above criterion as the CSYWO criterion, which is an alternative form of Peres-type criterion for steering. However, the paper of Ref. \cite{Chen2011} has never been submitted to any physical journal for publication, probably due to its lack of rigorous proof.

Nonetheless, Ref. \cite{Chen2011} has provided some inspirational clues on how to construct the Peres-type criterion for steering. In the following, let us briefly review the CSYWO criterion, and analyze its advantages and deficiency on detecting the steerability. The CSYWO criterion is given by
\begin{eqnarray}
\mathcal
{S}=\lambda_1+\lambda_2-(\lambda_1-\lambda_2)^2<0,
\label{CSYWO}
\end{eqnarray}
where $\lambda_1$ and $\lambda_2$ are respectively the smallest and the second small eigenvalues of $\rho_{AB}^{\rm T_B}$. Ref. \cite{Chen2011} mentioned that, when (\ref{CSYWO}) is satisfied, then the steerability exists in the two-qubit state $\rho_{AB}$.

\emph{Advantages.---} Let us analyze some advantages of the CSYWO criterion.

(i) The form of (\ref{CSYWO}) is very simple and has an aesthetic appeal. Its construction only depends on two eigenvalues $\lambda_1$ and $\lambda_2$. Although it is not symmetric under the permutation of the set $\{\lambda_1,\lambda_2,\lambda_3,\lambda_4\}$, it indeed has a permutation symmetry between $\lambda_1$ and $\lambda_2$.

(ii) It is well-known that quantum entanglement, EPR steering and Bell's nonlocality have a strict hierarchical structure \cite{WJD07,WJD07PRA}, i.e., EPR steering is a subset of quantum entanglement, and Bell's nonlocality is a subset of EPR steering. In other words, if the state $\rho_{AB}$ is separable, then it must not be steerable; if the state $\rho_{AB}$ is entangled, then it has an opportunity to be steerable, but is not necessary steerable. In a certain sense, the construction of (\ref{CSYWO}) reflects the above hierarchical property. Let us explain this point in detail as follows.

The criterion (\ref{CSYWO}) is a kind of generalization of Peres's criterion through adding an additional item, i.e.,
\begin{eqnarray}
\mathcal{S} & =& \lambda_1+\boxed{\lambda_2-(\lambda_1-\lambda_2)^2}\nonumber\\
& =& \mathcal{E}+\boxed{\lambda_2-(\lambda_1-\lambda_2)^2},
\label{CSYWO-1}
\end{eqnarray}
here the boxed term is just the additional item. Firstly, if the state $\rho_{AB}$ is separable, which means $\lambda_1$ and $\lambda_2$ are positive, then one automatically has $\mathcal{S}  = \lambda_1+\lambda_2-(\lambda_1-\lambda_2)^2=\lambda_1(1-\lambda_1) + \lambda_2(1-\lambda_2)+2\lambda_1\lambda_2 \geq 0$. Secondly, if the state $\rho_{AB}$ is entangled, this means that the additional item has a chance to be positive number, otherwise $\mathcal{E}<0$ directly leads to $\mathcal{S}<0$ (i.e., there will be no any gap between entanglement and steering). One may check that for some quantum states the additional item can be positive. For instance, for the two-qubit Werner state, we have $\lambda_1=(1-3V)/4$ and $\lambda_2=(1+V)/4$, then the additional item $\lambda_2-(\lambda_1-\lambda_2)^2=-V^2+(1+V)/4$ can be positive in the region $V\in (1/3, 1/2]$, where there is entanglement but without steering.

(iii) The criterion (\ref{CSYWO}) is a necessary and sufficient condition to detect all pure entangled states of two qubits. For example, let us consider the region $\theta\in [0, \pi/4]$, then from Eq. (\ref{eq:pure-1}) we have  $\lambda_1=-\sin\theta\cos\theta$ and $\lambda_2=\sin^2\theta$, thus
\begin{eqnarray}
\mathcal {S}=-\frac{1}{2}\sin2\theta(1+2\sin^2\theta)<0.
\end{eqnarray}
For the region $\theta\in [\pi/4, \pi/2]$, one obtains the similar result.

(iv) The criterion (\ref{CSYWO}) is a necessary and sufficient condition to detect the critical value $V_{\rm cr}^{\rm S}=1/2$ of steering for Werner state. From Eq. (\ref{eq:werner1}), we have $\lambda_1=(1-3V)/4$ and $\lambda_2=(1+V)/4$, thus the criterion
\begin{eqnarray}
\mathcal {S}=-\frac{1}{2}(2V-1)(1+V)<0
\end{eqnarray}
successfully detects the critical value $V_{\rm cr}^{\rm S}=1/2$ for EPR steering.

(v) The all-versus-nothing (AVN) state is presented in 2013 in Ref. \cite{2013AVN}, which is given by
{\small
\begin{eqnarray}\label{eq:avn0}
&&\rho_{\rm AVN} =\nonumber\\
&& \begin{bmatrix}
 \nu_1 \cos^2 \theta    & 0  &  0  & \nu_1 \sin \theta \cos \theta    \\
 0 &  \nu_2\sin ^2\theta  & \nu_2  \sin \theta \cos \theta  & 0 \\
 0 & \nu_2 \sin \theta \cos \theta  &  \nu_2\cos^2\theta   & 0 \\
 \nu_1 \sin \theta \cos \theta  & 0  &  0  &  \nu_1 \sin^2 \theta \\
\end{bmatrix}, \nonumber\\
\end{eqnarray}
}
\noindent with $\theta\in [0, \pi/2]$, $\nu_1 \in[0, 1]$, $\nu_2\in [0, 1]$ and $\nu_1+\nu_2=1$. The degree of the entanglement (i.e., the concurrence \cite{Wootters1998}) for the state $\rho_{\rm AVN}$ reads $\mathcal{C}=|\nu_1 - \nu_2| \sin2\theta$.
By applying the AVN proof of EPR steering, Ref. \cite{2013AVN} has proved that the state $\rho_{\rm AVN}$ is steerable in all the entangled region (i.e., $\theta \neq 0$ and $\pi/2$, and $\nu_1\neq \nu_2$). Remarkably, the criterion (\ref{CSYWO}) is a necessary and sufficient condition to detect the steerability of $\rho_{\rm AVN}$. Explicitly, in this situation, one has four eigenvalues of $\rho_{\rm AVN}^{\rm T_B}$ as
\begin{eqnarray}\label{eq:avn}
\lambda & \in & \left\{ \frac{1}{2} \bigl( \nu_1 \pm \sqrt{\nu_1^2 \cos^2 2\theta + \nu_2^2 \sin^2 2\theta} \bigr), \right. \nonumber \\
        && \left. \frac{1}{2} \bigl( \nu_2 \pm \sqrt{\nu_2^2 \cos^2 2\theta + \nu_1^2 \sin^2 2\theta} \bigr) \right\}.
\end{eqnarray}
Without loss of generality, we can assume $\nu_2\geq\nu_1$. From Eq. (\ref{eq:avn}), we have
\begin{eqnarray}
&& \lambda_1=\frac{1}{2} \left( \nu_1 - \sqrt{\nu_1^2 \cos^2 2\theta + \nu_2^2 \sin^2 2\theta} \right), \nonumber\\
&& \lambda_2=\frac{1}{2} \left( \nu_2 - \sqrt{\nu_2^2 \cos^2 2\theta + \nu_1^2 \sin^2 2\theta} \right),
\end{eqnarray}
and
\begin{eqnarray}
\lambda_1+\lambda_2 &=& \frac{1}{2} -
\frac{1}{2} \Big(\sqrt{\nu_1^2 \cos^2 2\theta + \nu_2^2 \sin^2 2\theta} \nonumber\\
&&+\sqrt{\nu_2^2 \cos^2 2\theta + \nu_1^2 \sin^2 2\theta}\Big).
\end{eqnarray}
With the help of
\begin{eqnarray}
&& \bigl(\nu_2^2 \cos^2 2\theta + \nu_1^2 \sin^2 2\theta\bigr)\bigl(\nu_1^2 \cos^2 2\theta + \nu_2^2 \sin^2 2\theta\bigr)\nonumber \\
&&=\nu_1^2 \nu_2^2+ (\nu_1 - \nu_2)^2 \sin^2 2\theta \cos^2 2\theta,
\end{eqnarray}
one can prove that
\begin{eqnarray}
&&\Big(\sqrt{\nu_1^2 \cos^2 2\theta + \nu_2^2 \sin^2 2\theta} +\sqrt{\nu_2^2 \cos^2 2\theta + \nu_1^2 \sin^2 2\theta}\Big)^2\nonumber\\
&&=\nu_1^2+\nu_2^2 +2 \sqrt{\nu_1^2 \nu_2^2+ (\nu_1 - \nu_2)^2 \sin^2 2\theta \cos^2 2\theta}\nonumber\\
&&\geq \nu_1^2+\nu_2^2 +2 \nu_1 \nu_2= (\nu_1+\nu_2)^2=1,
\end{eqnarray}
which yields
\begin{eqnarray}
\lambda_1+\lambda_2\leq0.
\end{eqnarray}
Thus $\mathcal{S}  = \lambda_1+\lambda_2-(\lambda_1-\lambda_2)^2\leq0$.
Here the equality only happens when $\lambda_1=\lambda_2=0$ (i.e. $\mathcal{C}=0$).

\emph{Deficiency.---} Let us analyze a deficiency of the CSYWO criterion. Besides the above advantages, the the CSYWO criterion is also applicable to detect the steerability of some important rank-2 states, such as the following specific Bell-diagonal state
\begin{eqnarray}
\rho
&=&\left(\begin{matrix}
\frac{V}{2}&0&0&\frac{V}{2}\\
0&\frac{1-V}{2}&\frac{1-V}{2}&0\\
0&\frac{1-V}{2}&\frac{1-V}{2}&0\\
\frac{V}{2}&0&0&\frac{V}{2}\end{matrix} \right),
\end{eqnarray}
and the non-maximal entangle state mixed with the color noise
\begin{eqnarray}
\rho
&=&\left(\begin{matrix}
V\cos^2\theta+\frac{1-V}{2}&0&0&V\sin\theta\cos\theta\\
0&0&0&0\\
0&0&0&0\\
V\sin\theta\cos\theta&0&0&V\sin^2\theta+\frac{1-V}{2}\end{matrix}
\right).
\end{eqnarray}
These two rank-2 states violate the Clauser-Horne-Shimony-Holt inequality \cite{1969CHSH}, hence possessing steerability. For the detail, the readers may refer to example 3 and example 4 in Ref. \cite{Chen2011}.

Very recently, a Gisin-like fundamental theorem has been established for EPR steering, which indicates all rank-2 (and rank-1) entangled states possess EPR steerability. Thus all rank-2 entangled states can be applicable as EPR-steering resources in quantum information. Now, a deficiency of the CSYWO criterion appears -- it is not able to detect the steerability of some rank-2 entangled states. For example, for the following rank-2 state
\begin{eqnarray}
\rho=\frac{1}{2}  (\ket{\psi_1}\bra{\psi_1}+ \ket{\psi_2}\bra{\psi_2}),
\end{eqnarray}
where
\begin{eqnarray}
&&\ket{\psi_1} = \frac{1}{\sqrt{2}}(\ket{00}+\ket{11}), \nonumber\\
&& \ket{\psi_2} = \ket{01}.
\end{eqnarray}
The concurrence of the state $\rho$ is given by $\mathcal{C}=1/2$. According to the theorem in Ref. \cite{Chen2025}, the state $\rho$ is steerable.
The four eigenstates of $\rho^{\rm T_B}$ reads
\begin{eqnarray}
&&\lambda_1=\frac{1-\sqrt{2}}{4}, \;\;\lambda_2= \lambda_3=\frac{1}{4}, \;\;\lambda_4=\frac{1+\sqrt{2}}{4},
\end{eqnarray}
which yields
\begin{eqnarray}
\mathcal{S}&=& \lambda_1+\lambda_2-(\lambda_1-\lambda_2)^2 =\frac{3}{8}-\frac{\sqrt{2}}{4}>0.
\end{eqnarray}
Thus the CSYWO criterion fails to detect the steerability of the specific rank-2 state $\rho$.

This motivates us to develop a more efficient Peres-type criterion for EPR steering. At least, the new Peres-type criterion serves as a necessary and sufficient condition to detect the steerability for the following three family states: (i) The two-qubit Werner state; (ii) All two-qubit pure (i.e., rank-1) states; (iii) All two-qubit rank-2 states.

\section{Construction of a New Peres-type Criterion}\label{sec:peres3}

In this section, let us come to construct a more efficient Peres-type criterion for EPR steering. The key point is how to construct the quantity $\mathcal{S}$ involved in the Peres-type criterion.

Similar to Eq. (\ref{CSYWO-1}), the quantity $\mathcal{S}$ can be written as
\begin{eqnarray}
\mathcal{S} & =& \mathcal{E}+\boxed{{\rm Additional \;\; Item}}.
\label{new-1}
\end{eqnarray}
However, here the quantity $\mathcal{E}$ is taken as
\begin{eqnarray}
\mathcal {E}= \Lambda_4 =\lambda_1 \lambda_2 \lambda_3 \lambda_4,
\end{eqnarray}
which has a permutation symmetry among the set $\{\lambda_1,\lambda_2,\lambda_3,\lambda_4\}$. Of course, one may multiply a constant positive number $k$ on the right-hand-side of Eq. (\ref{new-1}), i.e.,
\begin{eqnarray}
\mathcal{S} & =& k \, \left(\mathcal{E}+\boxed{{\rm Additional \;\; Item}}\; \right).
\label{new-2}
\end{eqnarray}
as expect, this would not affect the criterion. Since symmetry plays a crucial role in physics, thus we expect that the additional term also has a
permutation symmetry among the set $\{\lambda_1,\lambda_2,\lambda_3,\lambda_4\}$. Then as a whole, the quantity $\mathcal{S}$  has a permutation symmetry. As a result, one can have
\begin{eqnarray}
{\rm Additional \;\; Item} = F (\Lambda_1, \Lambda_2, \Lambda_3, \Lambda_4),
\end{eqnarray}
where $F$ is a function of $\Lambda_j$'s.

Now we come to study the underlying symmetry and the underlying constraint between these $\Lambda_j$'s. Let us start from the normalization condition
\begin{eqnarray}\label{eq:Lambda1}
&& \Lambda_1=\lambda_1+\lambda_2+\lambda_3+\lambda_4=1,
\end{eqnarray}
which corresponds ${\rm tr} [\rho^{\rm T_B}]={\rm tr} \rho=1$. Based on Eq. (\ref{eq:Lambda1}), we have
\begin{eqnarray}
&& (\lambda_1+\lambda_2+\lambda_3+\lambda_4)^2=1,
\end{eqnarray}
i.e.,
\begin{eqnarray}
&& \lambda_1^2+\lambda_2^2+\lambda_3^2+\lambda_4^2 + 2 \sum\limits_{1 \leq i < j \leq 4} (\lambda_i \lambda_j )=1,
\end{eqnarray}
i.e.,
\begin{eqnarray}
&& \frac{1}{3}\sum\limits_{1 \leq i < j \leq 4}(\lambda_i^2+\lambda_j^2)+ 2 \sum\limits_{1 \leq i < j \leq 4} (\lambda_i \lambda_j )=1.
\end{eqnarray}
By using $x^2+y^2 \geq 2 xy$, we obtain
\begin{eqnarray}
&& \frac{1}{3}\times  2 \sum\limits_{1 \leq i < j \leq 4} (\lambda_i \lambda_j )+ 2 \sum\limits_{1 \leq i < j \leq 4} (\lambda_i \lambda_j )\leq 1,
\end{eqnarray}
which leads to
\begin{eqnarray}\label{eq:constr-0}
&& \Lambda_2=\sum\limits_{1 \leq i < j \leq 4} (\lambda_i \lambda_j )\leq \frac{3}{8}.
\end{eqnarray}

Next, let us study the square of $\Lambda_3$, i.e.,
\begin{eqnarray}
&&\Lambda_3^2 =(\sum\limits_{1 \leq i < j < k \leq 4} \lambda_i \lambda_j \lambda_k)^2.
\end{eqnarray}
For convenience, let us denote
\begin{eqnarray}
&&a_l=\lambda_i \lambda_j \lambda_k,
\end{eqnarray}
with $l \neq i, j, k$, for example, $a_1=\lambda_2 \lambda_3 \lambda_4$. Then
\begin{eqnarray}
&&\Lambda_3^2 =(\sum\limits_{1 \leq l \leq 4} a_l)^2.
\end{eqnarray}
Let us start from the following inequality
\begin{eqnarray}
&&\sum\limits_{1 \leq l<m \leq 4} (a_l- a_m)^2\geq0.
\end{eqnarray}
One has
\begin{eqnarray}
3\sum\limits_{1 \leq l \leq 4} a_l^2\geq2\sum\limits_{1 \leq l< m \leq 4} a_l a_m,
\end{eqnarray}
i.e.,
\begin{eqnarray}
\sum\limits_{1 \leq l \leq 4} a_l^2\geq\frac{2}{3}\sum\limits_{1 \leq l< m \leq 4} a_l a_m.
\end{eqnarray}
Then we have
\begin{eqnarray}
\Lambda_3^2 &=& (\sum\limits_{1 \leq l \leq 4} a_l)^2 \nonumber\\
&=& \sum\limits_{1 \leq l \leq 4} a_l^2+2\sum\limits_{1 \leq l< m \leq 4}a_l a_m\nonumber\\
&\geq & \frac{8}{3}\sum\limits_{1 \leq l< m \leq 4} a_l a_m,
\end{eqnarray}
which is
\begin{eqnarray}
&&(\sum\limits_{1 \leq i < j < k \leq 4} \lambda_i \lambda_j \lambda_k)^2\geq\frac{8}{3}(\lambda_1\lambda_2\lambda_3\lambda_4)\sum\limits_{1 \leq i < j \leq 4} (\lambda_i \lambda_j). \nonumber\\
\end{eqnarray}
i.e.,
\begin{eqnarray}\label{eq:constr}
&& \Lambda_3^2 \geq\frac{8}{3} \Lambda_4 \Lambda_2.
\end{eqnarray}
This is the underlying constraint between $\Lambda_j$'s. Thus we can write the additional item as
\begin{eqnarray}
F (\Lambda_1, \Lambda_2, \Lambda_3, \Lambda_4)=k_2 \Lambda_4 \Lambda_2+ k_3 \Lambda_3^2,
\end{eqnarray}
in turn, the quantity $\mathcal{S}$ can be written as
\begin{eqnarray}
\mathcal{S} & =& k_1 \mathcal{E}+ F (\Lambda_1, \Lambda_2, \Lambda_3, \Lambda_4) \nonumber\\
&=& k_1 \Lambda_4 + k_2 \Lambda_4 \Lambda_2+ k_3 \Lambda_3^2,
\end{eqnarray}
where $k_1$, $k_2$ and $k_3$ are some real numbers. Without loss of generality, we have
\begin{eqnarray}
\mathcal{S}
&=& (1+\mu) \Lambda_4 (1 + \tau \Lambda_2)+ \Lambda_3^2,
\end{eqnarray}
where $\mu$ and $\tau$ are real numbers with $(1+\mu)\neq 0$, and for simplicity the coefficient of $\Lambda_3^2$ has been normalized to be 1.

The Peres-type criterion is given by
\begin{eqnarray}\label{eq:newcrite-1}
\mathcal{S}
&=& (1+\mu) \Lambda_4 (1 + \tau \Lambda_2)+ \Lambda_3^2 <0.
\end{eqnarray}
if (\ref{eq:newcrite-1}) is satisfied, then the two-qubit state $\rho_{AB}$ possesses steerability. In the following, let us come to determine the parameters $\mu$ and $\tau$.

(i) We would like to require the criterion (\ref{eq:newcrite-1}) is a necessary and sufficient condition to detect the two-qubit Werner state. It has been known that the critical value of steerability is $V_{\rm cr}^{\rm S}=1/2$, i.e., the steerable region is $V\in (1/2, 1]$ and the unsteerable region is $V\in [0, 1/2]$. From Eq. (\ref{eq:werner1}), we have
\begin{eqnarray}\label{eq:lambda-1a}
&& \lambda_1=\frac{1-3V}{4}, \;\; \lambda_2=\lambda_3=\lambda_4=\frac{1+V}{4}.
\end{eqnarray}
For the critical point $V=1/2$, we should have $\mathcal{S}=0$.
Because
\begin{eqnarray}
\Lambda_3^2|_{V=1/2}=0,
\end{eqnarray}
then we have
\begin{eqnarray}
(1 + \tau \Lambda_2)|_{V=1/2}=0,
\end{eqnarray}
i.e.,
\begin{eqnarray}
\left(1 + 3 \tau \frac{1-V^2}{8}\right)\big|_{V=1/2}=0,
\end{eqnarray}
i.e.,
\begin{eqnarray}
1 + \frac{9\tau}{32}=0.
\end{eqnarray}
Then we have the parameter
\begin{eqnarray}
\tau=- \frac{32}{9}=0.
\end{eqnarray}
In this situation, we have the quantity $\mathcal{S}$ as
\begin{eqnarray}\label{eq:SS-1}
\mathcal{S}
&=& (1+\mu) \Lambda_4 \left(1 - \frac{32}{9}\Lambda_2\right)+ \Lambda_3^2.
\end{eqnarray}
Substitute Eq. (\ref{eq:lambda-1a}) into Eq. (\ref{eq:SS-1}), we have
\begin{eqnarray}
\mathcal{S}
&=& (1+\mu) \lambda_1 \lambda_2^3 \left[1 - \frac{32}{3} \lambda_2(\lambda_1+\lambda_2)\right] \nonumber\\
&&+ \left[\lambda_2^2(3\lambda_1+\lambda_2)\right]^2,
\end{eqnarray}
i.e.,
\begin{eqnarray}
\mathcal{S}&=& (1+\mu) \lambda_1 \lambda_2^3 \left[1 - \frac{4}{3} (1-V^2)\right] + \lambda_2^4(1-2V)^2\nonumber\\
&=& -\frac{1}{3}(1+\mu) \lambda_1 \lambda_2^3 (1-4V^2)+ \lambda_2^4(1-2V)^2\nonumber\\
&=& \lambda_2^3(1-2V) \left[-\frac{1}{3}(1+\mu) \lambda_1 (1+2V)+ \lambda_2(1-2V)\right]\nonumber\\
&=& \frac{\lambda_2^3(1-2V)}{12} \times \nonumber\\
&&\left[-(1+\mu) (1-3V) (1+2V)+ 3(1+V)(1-2V)\right]\nonumber\\
&=& \frac{\lambda_2^3(1-2V)}{12} \left[6 \mu V^2 +(\mu-2)V+ (2-\mu)\right]
\nonumber\\
&=& -\frac{\lambda_2^3(2V-1)}{12} \left[6 \mu V^2 +(\mu-2)V+ (2-\mu)\right].
\end{eqnarray}
Because $\mathcal{S} < 0$ for $V\in (1/2, 1]$, and $\mathcal{S} \geq 0$ for $V\in [0, 1/2]$, then we have the region of the parameter $\mu$ as
\begin{eqnarray}
\mu\in (0, 2].
\end{eqnarray}

(ii) Let us verify whether the criterion (\ref{eq:newcrite-1}) is a necessary and sufficient condition to detect all two-qubit pure entangled states. Based on Eq. (\ref{eq:pure-1}), we have
\begin{eqnarray}
&& \Lambda_2=0, \nonumber\\
&& \Lambda_3=-\sin^2\theta  \cos^2\theta, \nonumber\\
&& \Lambda_4=-\sin^4\theta  \cos^4\theta,
\end{eqnarray}
which leads to
\begin{eqnarray}
\mathcal{S}
&=& -\mu \sin^4\theta  \cos^4\theta.
\end{eqnarray}
Because $0<\mu \leq 2$, thus $\mathcal{S}<0$ for all the entangled region $\theta\in (0, \pi/2)$. This ends the verification.

\begin{remark}As a criterion, $\mathcal{S}$ should be non-negative for all separable states. Thus we need further to prove that
\begin{eqnarray}\label{eq:SS-1a}
\mathcal{S}
&=& (1+\mu) \Lambda_4 \left(1 - \frac{32}{9}\Lambda_2\right)+ \Lambda_3^2\geq 0
\end{eqnarray}
holds for any $\lambda_j \geq 0$ $(j=1, 2, 3, 4)$. Because $\lambda_j\geq 0$, thus all $\Lambda_j$'s are non-negative. Due to the constraint (\ref{eq:constr}), we have
\begin{eqnarray}
\mathcal{S}
&=& (1+\mu) \Lambda_4 - \frac{32}{9}(1+\mu) \Lambda_4 \Lambda_2+ \Lambda_3^2 \nonumber\\
&\geq & (1+\mu) \Lambda_4 - \frac{32}{9}(1+\mu) \Lambda_4 \Lambda_2+ \frac{8}{3} \Lambda_4 \Lambda_2 \nonumber\\
&\geq & (1+\mu) \Lambda_4 +\frac{8}{3} \Lambda_4 \Lambda_2 \left[1- \frac{4}{3}(1+\mu)\right].
\end{eqnarray}
Because $\mu\in(0, 2]$, then $[1- (4/3)(1+\mu)]<0$, due to the constraint (\ref{eq:constr-0}), the above equation becomes
\begin{eqnarray}
\mathcal{S}
&\geq & (1+\mu) \Lambda_4 +\frac{8}{3} \Lambda_4 \Lambda_2 \left[1- \frac{4}{3}(1+\mu)\right]\nonumber\\
&\geq & (1+\mu) \Lambda_4 + \Lambda_4  \left[1- \frac{4}{3}(1+\mu)\right]\nonumber\\
&= & \frac{2-\mu}{3} \Lambda_4 \geq 0.
\end{eqnarray}
This ends the proof. $\blacksquare$
\end{remark}

Up to now, we have established a Peres-type criterion of EPR steering, which is a necessary and sufficient condition for (i) the two-qubit Werner state and (ii) all two-qubit pure (i.e., rank-1) states. In the next section, we shall prove that it is also a necessary and sufficient condition for detecting the steerability of all two-qubit rank-2 states.

%
%

\section{The Proof for All Two-Qubit Rank-2 States}\label{sec:peres4}

As shown in \cite{Chen2025}, under the local unitary transformations, for a two-qubit density matrix with rank 2, it is sufficient to consider the following form
\begin{eqnarray}
\rho=\nu_1  \ket{\psi_1}\bra{\psi_1}+\nu_2 \ket{\psi_2}\bra{\psi_2},
\end{eqnarray}
where
\begin{eqnarray}
\ket{\psi_1} &=& \cos \theta \ket{00}+\sin \theta\ket{11}, \nonumber\\
 \ket{\psi_2} &=& \cos\phi(\cos \alpha \ket{01} +\sin\alpha\ket{10})\nonumber\\
 && + e^{{\rm i}\beta}\sin\phi(\sin \theta \ket{00}-\cos \theta\ket{11}),
\end{eqnarray}
with
\begin{eqnarray}
&& {\theta\in [0, \pi/2]}, \;\; {\phi \in [0, \pi/2]}, \;\;  {\alpha \in [0, \pi/2]}, \;\; \beta \in [0, 2\pi],   \nonumber\\
&& \nu_1 \in [0,1], \;\;\; \nu_2 \in [0,1], \;\;\; \nu_1+\nu_2=1.
\end{eqnarray}
Here $| \psi_1\rangle$  and $|\psi_2\rangle$ are two orthogonal pure states of two qubits, i.e., $\langle \psi_1| \psi_2\rangle =\langle \psi_2| \psi_1\rangle=0$, and the parameters $\nu_i$ ($i=1,2$) are the weights of pure states $|\psi_i\rangle$ ($i=1,2$) in $\rho$ respectively. For a true rank-2 state, one generally has $\nu_1\neq \nu_2$.

Let us study the eign-equation of $\rho^{T_B} $. From
\begin{eqnarray}
\det[\lambda\, \mathbb{I}-\rho^{T_B} ]=0,
\end{eqnarray}
one can have
\begin{eqnarray}\label{eq:lambda-1b}
\lambda^4 - \lambda^3 + \nu_1\nu_ 2\lambda^2 + (\nu_1 - \nu_2)\Omega\, \lambda+ \Gamma-\Omega^2=0,
\end{eqnarray}
with
\begin{eqnarray}
\Omega &=& \Omega_1-\Omega_2, \nonumber\\
\Gamma &=& \nu_1\nu_2\cos^22\theta\sin^2\phi(\Gamma_1  - \Omega_1 -\Omega_2),
\end{eqnarray}
and
\begin{eqnarray}
 \Gamma_1&=& \nu_1\nu_2\cos\theta\sin\theta\big(\cos2\beta\cos^2\phi\sin2\alpha + \sin2\theta\sin^2\phi\big), \nonumber\\
\Omega_1 &=&\nu_1^2\cos^2\theta\sin^2\theta, \nonumber\\
 \Omega_2&=& \nu_2^2\big(\cos^2\alpha\cos^4\phi\sin^2\alpha  + \cos^2\theta\sin^2\theta\sin^4\phi\nonumber\\
 &&+ 2\cos\alpha\cos2\beta\cos\theta\cos^2\phi\sin\alpha\sin\theta\sin^2\phi\big).
\end{eqnarray}
Based on Eq. (\ref{eq:lambda-1b}), we can obtain
\begin{eqnarray}
&& \Lambda_1=1, \nonumber\\
&& \Lambda_2=\nu_1\nu_2, \nonumber\\
&& \Lambda_3=-(\nu_1 - \nu_2)\Omega, \nonumber\\
&& \Lambda_4=\Gamma-\Omega^2.
\end{eqnarray}

\begin{remark}
\emph{The quantity $\Gamma$ is always non-positive.} Here, we would like to prove that
\begin{eqnarray}
\Gamma \leq 0.
 \end{eqnarray}
Because $\Omega_1\geq 0$ and $\Omega_2\geq 0$, we can have
\begin{eqnarray}
\Gamma &=& \nu_1\nu_2\cos^22\theta\sin^2\phi(\Gamma_1  - \Omega_1 -\Omega_2)\nonumber\\
&\leq & \nu_1\nu_2\cos^22\theta\sin^2\phi(\Gamma_1  - 2 \sqrt{\Omega_1\Omega_2}).
\end{eqnarray}
Due to
\begin{eqnarray}
&&\Gamma_1^2  - (2 \sqrt{\Omega_1\Omega_2})^2 \nonumber\\
&=& -\nu_1\nu_2\cos\theta\sin\theta \cos^4\phi \sin^22\alpha\sin^22\beta\leq 0,
\end{eqnarray}
and $\nu_1>0$, $\nu_2>0$, $\theta\in [0, \pi/2]$, we immediately have $\Gamma \leq 0$. It ends the proof. $\blacksquare$
\end{remark}

\begin{remark} \emph{The relation between $(\Gamma-\Omega^2)$ and $\mathcal{C}^2$.} According to Ref. \cite{Wootters1998}, the concurrence of the state $\rho$ is given by
\begin{eqnarray}
\mathcal{C} &=& \sqrt{-s_2-2\sqrt{s_1}},
\end{eqnarray}
where
\begin{eqnarray}\label{eq:s-1}
 s_2 &=& -\frac{1}{2} \nu_2^2 \sin2 \alpha  \sin 2 \theta \sin ^2 2 \phi \cos 2 \beta-\nu_2^2 \sin ^2 2 \alpha \cos ^4 \phi\nonumber\\
  &&-\sin ^2 2 \theta \left(\nu_1^2-\nu_1 \nu_2 \sin ^2\phi +\nu_2^2 \sin ^4\phi \right)
\nonumber\\
&&-\frac{1}{2} \nu_1 \nu_2 (\cos4 \theta+3) \sin ^2 \phi, \nonumber\\
 s_1 &=& \frac{1}{2} \nu_1^2 \nu_2^2 (\sin 2 \alpha \sin2 \theta \sin ^2 2 \phi \cos 2 \beta \nonumber\\
 &&+2 \sin ^2 2 \alpha \sin ^2 2 \theta \cos ^4\phi +2 \sin ^4\phi ),
\end{eqnarray}
with $-s_2\geq 0$ and $s_1>0$. Interestingly, after some careful calculations we find that
\begin{eqnarray}
16(\Gamma-\Omega^2) + (s_2^2-4s_1)=0,
\end{eqnarray}
i.e.,
\begin{eqnarray}\label{eq:goc}
\Gamma-\Omega^2 &=& -\frac{1}{16} (-s_2+2 \sqrt{s_1})\, \mathcal{C}^2.
\end{eqnarray}
Note that $-s_2 \geq 0$ and $(-s_2+2 \sqrt{s_1}) \geq 0$, thus $\Gamma-\Omega^2 \leq 0$.

(i) The above result coincides with Peres's criterion for entanglement. Because $\Lambda_4=\Gamma-\Omega^2$, thus if the state is entangled (i.e., $\mathcal{C}\neq 0$), one direct has $\mathcal{E}=\Lambda_4 <0$.

(ii) Because $\Gamma\leq 0$, based on Eq. (\ref{eq:goc}), if $\mathcal{C}=0$, then one must have $\Gamma=0$ and $\Omega=0$, and vice versa.
$\blacksquare$
\end{remark}

Now we are ready to prove the new Peres-type criterion is also a necessary and sufficient condition for detecting the steerability of all two-qubit rank-2 states. In this situation, the quantity $\mathcal{S}$ is given by
\begin{eqnarray}\label{eq:ss-1b}
\mathcal{S}
&=& (1+\mu) \Lambda_4 \left(1 - \frac{32}{9}\Lambda_2\right)+ \Lambda_3^2 \nonumber\\
&=& (1+\mu) (\Gamma-\Omega^2) \left(1 - \frac{32}{9}\nu_1\nu_2 \right)+ (\nu_1 - \nu_2)^2\Omega^2 \nonumber\\
&=& (1+\mu) \left(1 - \frac{32}{9}\nu_1\nu_2 \right)\times \nonumber\\
&&\left[\Gamma-\Omega^2 + \frac{(\nu_1 - \nu_2)^2}{(1+\mu) \left(1 - \frac{32}{9}\nu_1\nu_2 \right)}\Omega^2 \right].
\end{eqnarray}
Because $\nu_1\nu_2\leq 1/4$, $(\nu_1 - \nu_2)^2=1-4\nu_1\nu_2$, $0<\nu\leq 2$, then we have
\begin{eqnarray}
&&1-\frac{32}{9}\nu_1\nu_2>0\nonumber\\
&&(1+\mu)(1-\frac{32}{9}\nu_1\nu_2)>(1-4\nu_1\nu_2).
\end{eqnarray}
Furthermore, due to $\Gamma\leq 0$ and $\Omega^2\geq 0$, from Eq. (\ref{eq:ss-1b}) we immediately have
we can get $\mathcal{S}\leq0$, and the equality only happens when $\Gamma=0$ and $\Omega=0$. This ends the proof.

\section{Examples for Rank-3 and Rank-4 states} \label{sec:peres5}

In this section, let us check the Peres-type criterion by two well-known rank-3 and rank-4 states. We find that the results are reasonable.

\emph{Example 1.---} \emph{The maximally entangled mixed state (MEMS) }\cite{MEMS2001}. The MEMS is given by
\begin{eqnarray}
\rho_{\rm MEMS}=\left(\begin{matrix}
g(\gamma)&0&0&\gamma/2\\
0&1-2g(\gamma)&0&0\\
0&0&0&0\\
 \gamma/2&0&0&g(\gamma)\end{matrix} \right),
\end{eqnarray}
with $g(\gamma)=1/3$ for $\gamma\in[0,2/3]$ and $g(\gamma)=\gamma/2$
for $\gamma\in[2/3,1]$. It violates the 10-setting steering
inequality in Ref.~\cite{NP2010} for $\gamma\geq 0.6029$. And it violates the 20-setting steering
inequality in Ref.~\cite{Chen2024} for $\gamma\geq 0.5806$.

In the case of $\gamma\in[0,2/3]$, the four eigenvalues of $\rho_{\rm MEMS}^{\rm T_B}$ read
\begin{eqnarray}
\lambda \in \left\{\frac{1-\sqrt{1+9\gamma^2}}{6}, \frac{1}{3}, \frac{1}{3}, \frac{1+\sqrt{1+9\gamma^2}}{6}\right\}.
\end{eqnarray}
Based on it, we can have
\begin{eqnarray}
\mathcal{S}=\frac{4 + 3(-7 + 5 \mu) \gamma^{2} + (9 - 72\mu ) \gamma^{4}}{2916}.
\end{eqnarray}

(i) For $\mu=0$, from $\mathcal{S}=0$, one obtains $\gamma=\gamma_1=\frac{\sqrt{33}-3}{6} \approx 0.457427$, i.e., $\mathcal{S}<0$ when $\gamma>\gamma_1$.

(ii) For $\mu=2$, from $\mathcal{S}=0$, one obtains $\gamma=\gamma_2=\frac{1}{3} \sqrt{\frac{1}{10} ( 3 + \sqrt{249} )} \approx 0.456797$, i.e.,
$\mathcal{S}<0$ when $\gamma>\gamma_2$.

Let us denote $\gamma_{\rm cr}$ be the critical value between EPR steering and the LHS model, then the above result suggests that $\gamma_{\rm cr} \in (0.456797, 0.457427)$. This result is reasonable, for it does not contradict with those obtained from the 10- and 20-setting steering inequalities. By the way, if one has a chance to know the exact number of $\gamma_{\rm cr}$ by other steering criterions, then inversely he may use it to determine the value of $\mu$.

\emph{Example 2.---} \emph{The generalized two-qubit Werner state.}
In Ref. \cite{Zukowski2014}, the researchers have developed a geometry-based steering inequality to study the generalized Werner state, which is given by
\begin{eqnarray}
\rho_{GW}=V |\psi_{\alpha} \rangle \langle \psi_{\alpha}| + (1 - V) \frac{\mathbb{I}}{4}
\end{eqnarray}
with $|\psi_{\alpha} \rangle=\sin(\alpha/2)|01 \rangle-\cos(\alpha/2)|10 \rangle$, $V\in[0,1]$, and $\alpha\in[0,\pi/2]$. Ref. \cite{Zukowski2014} has also proved a sufficient criterion for the steerablity, which is
\begin{eqnarray}\label{eq:marek}
V>\frac{3}{2(1+2\sin^2\alpha)}.
\end{eqnarray}
For $\alpha=\pi/2$, the state $\rho_{GW}$ reduces to the standard Werner state, and $V$ naturally recovers the critical value $V_{\rm cr}^{\rm S}=1/2$.

In the following, we would like compare the result obtained by our new Peres-type criterion $\mathcal{S}<0$ with that given in Ref. \cite{Zukowski2014}. Before it, we need to do some necessary calculations. The density matrix of the generalized Werner state is given by
\begin{eqnarray}
\rho_{GW}=
{\small
\begin{pmatrix}
 \frac{1-V}{4} & 0 & 0 & 0 \\
 0 & \frac{1-V}{4} + V \sin^2 \frac{\alpha}{2} & -V \cos \frac{\alpha}{2} \sin \frac{\alpha}{2} & 0 \\
 0 & -V \cos \frac{\alpha}{2} \sin \frac{\alpha}{2} & \frac{1-V}{4} + V \cos^2 \frac{\alpha}{2} & 0 \\
 0 & 0 & 0 & \frac{1-V}{4}
\end{pmatrix}. }\nonumber\\
\end{eqnarray}
Let us study the eigen-equation of $\rho_{GW}^{T_B} $. From
\begin{eqnarray}
\det[\lambda\, \mathbb{I}-\rho^{T_B} ]=0,
\end{eqnarray}
one can have
\begin{eqnarray}\label{eq.lambda3}
&&\lambda^4 - \lambda^3 + \frac{3}{8}(1 - V^2)\lambda^2 - \frac{1}{16}(1 -3V^2 + 2V^3 \cos 2\alpha)\lambda \nonumber\\
&& + \frac{1}{256}\bigl[1 - 6V^2 + V^4 + 4V^3 \cos 2\alpha (2 - V \cos 2\alpha)\bigr]\nonumber\\
&&=0.
\end{eqnarray}
Based on Eq. (\ref{eq.lambda3}), we can obtain
\begin{eqnarray}
&& \Lambda_1=1, \nonumber\\
&& \Lambda_2= \frac{3}{8}(1- V^2), \nonumber\\
&& \Lambda_3=\frac{1}{16}(1 - 3V^2 + 2V^3 \cos 2\alpha), \nonumber\\
&& \Lambda_4= \frac{1}{256}\bigl[1 - 6V^2 + V^4 + 4V^3 \cos 2\alpha (2- V \cos 2\alpha)\bigr].\nonumber\\
\end{eqnarray}
In this situation, the Peres-type criterion $\mathcal{S}$ is given by
\begin{eqnarray}\label{eq:crite}
\mathcal{S}
&=& (1+\mu) \Lambda_4 \left(1 - \frac{32}{9}\Lambda_2\right)+ \Lambda_3^2<0
\end{eqnarray}
with $\mu \in (0, 2]$. Now, let us make a comparison between the criterion (\ref{eq:marek}) and the Peres-type criterion (\ref{eq:crite}).

 (i) Let us consider the limit $V\to1$ and see what will happen. We can find $\rho_{GW}$ tends to a pure state. As a result, it is required to be steerable unless $\sin\alpha=0$. However, due to $V\leq 1$,  the criterion (\ref{eq:marek}) fails to detect the steerability of $\rho_{GW}$ for the region $\alpha\in [0, \pi/6]$. For the criterion (\ref{eq:crite}), when $V\to1$ we can have
\begin{eqnarray}
&&\mathcal{S}\to -\frac{\mu}{16} \sin^4 \alpha\leq 0.
\end{eqnarray}
This meets our expectation that it can detect the steerability for the whole entangled region. 

\begin{figure}[t]
    \includegraphics[width=75mm]{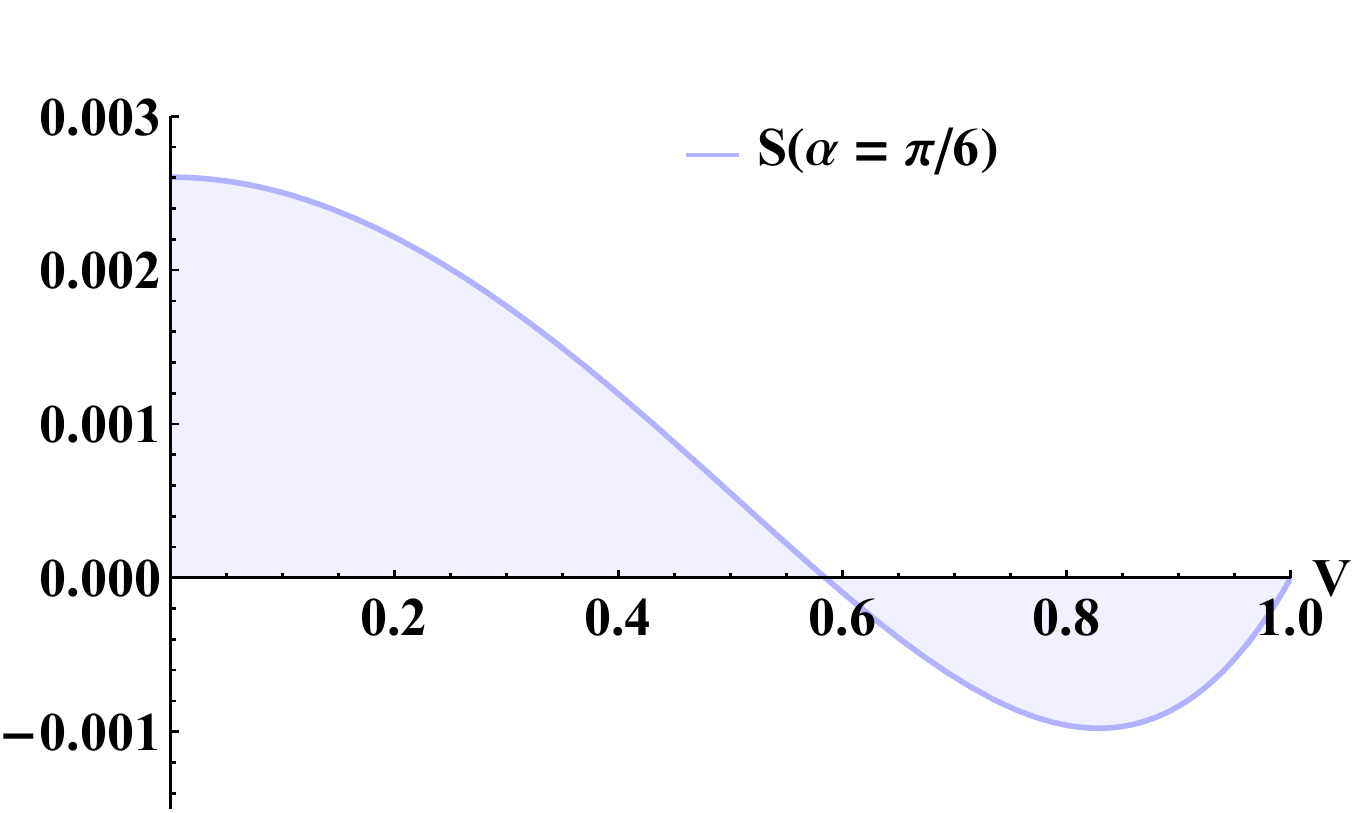}
    \caption{{\bf $\mathcal{S}$ versus $V$ for the fixed value $\alpha=\pi/6$}. Here $\mu \approx 0$. For $\alpha=\pi/6$, the criterion (\ref{eq:marek}) fails to detect the steerability of $\rho_{GW}$, however,  the criterion (\ref{eq:crite}) can detect the region of $V\in (0.5844, 1)$.
    }\label{fig:example1}
 \end{figure}

\begin{figure}[t]
    \includegraphics[width=75mm]{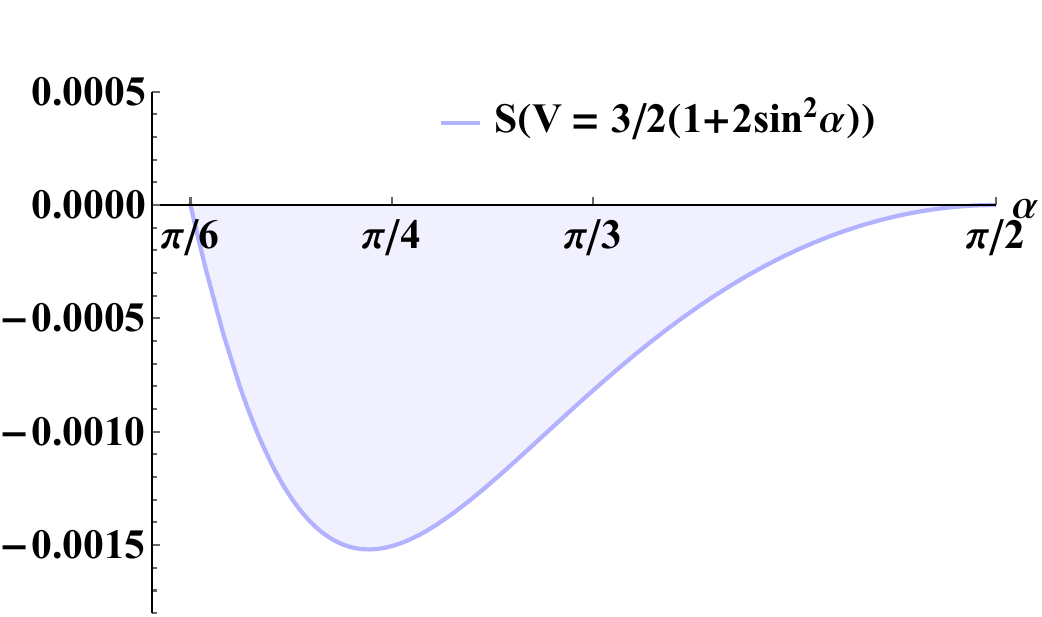}
    \caption{{\bf $\mathcal{S}$ versus $\alpha$ for $V=3/[2(1+2\sin^2\alpha)]$}. Here $\mu \approx 0$. For $V=3/[2(1+2\sin^2\alpha)]$, the criterion (\ref{eq:marek}) fails to detect the steerability of $\rho_{GW}$, however, the criterion (\ref{eq:crite}) can still detect the steerability.
    }\label{fig:example2}
 \end{figure}

(ii) Let us consider other $V$ and $\alpha$. One may observe that: (a) For $\alpha=\pi/6$, the criterion (\ref{eq:marek}) fails to detect the steerability of $\rho_{GW}$, however,  the criterion (\ref{eq:crite}) can detect the region of $V\in (0.5844, 1)$ (see in Fig. \ref{fig:example1}). (b) For $V=3/[2(1+2\sin^2\alpha)]$, the criterion (\ref{eq:marek}) fails to detect the steerability of $\rho_{GW}$, however, the criterion (\ref{eq:crite}) can still detect the steerability (see in Fig. \ref{fig:example2}).

\section{Conclusion and Outlook}

In conclusion, we have systematically investigated the possibility of constructing a Peres-type criterion for detecting EPR steering in two-qubit systems. Inspired by the simplicity and effectiveness of Peres's PPT criterion for entanglement, we aimed to develop an analogous criterion for steering that retains algebraic clarity and broad applicability.

We first reviewed the original Peres criterion and an earlier alternative (the CSYWO criterion) proposed for steering, analyzing its advantages and limitations. While the CSYWO criterion successfully detects steering in several important families of states -- such as pure states, Werner states, and certain rank-2 states -- it fails for some entangled rank-2 states, indicating the need for a more general and reliable criterion.

To address this, we constructed a new Peres-type steering criterion of the form (\ref{eq:crite}), which exhibits a permutation symmetry among the eigenvalues, reflecting the underlying structural symmetry of the steering problem. We rigorously proved that this criterion serves as a necessary and sufficient condition for detecting steering in: (i) the two-qubit Werner state;
(ii) all two-qubit pure states; and (iii) all two-qubit rank-2 states. Furthermore, the criterion naturally respects the hierarchical relationship among entanglement, steering, and Bell's nonlocality: it is satisfied only for states that are already entangled, and it correctly identifies the steering thresholds in known cases.

We also tested the criterion on higher-rank states, such as the maximally entangled mixed state and the generalized Werner state, and found that it yields reasonable steering thresholds consistent with existing inequality-based approaches. The criterion not only agrees with known results but in some cases predicts a broader steerable region, suggesting its potential usefulness in practical detection.

This work provides a significant step toward a unified spectral framework for detecting different forms of quantum nonlocality via partial transposition. The introduced criterion combines analytical tractability with broad applicability, offering a tool that is both theoretically elegant and practically useful.

Eventually, let us make some outlooks. Several directions remain for future research: (i) Determining the optimal value of the parameter $\mu$ through comparison with exactly solvable models or numerical optimization; (ii) Extending the criterion to higher-dimensional systems or multipartite scenarios; (iii) Providing a rigorous theoretical derivation of the criterion from first principles; (iv) Experimental verification using quantum optical or other physical platforms. Our results open a promising pathway for exploring Peres-like criteria beyond entanglement, enriching the toolbox for characterizing and harnessing quantum correlations in quantum information processing.

\begin{acknowledgments}
This work is supported by the Quantum Science and Technology-National Science and Technology Major Project (Grant No. 2024ZD0301000), and the National Natural Science Foundation of China (Grant No. 12275136).
\end{acknowledgments}


\begin{thebibliography}{99}
\bibitem{check1} Our numerical calculations have comfirmed this point. However, we have not found the relevent references that mentioned this point.
\end{thebibliography}
\end{document}